\documentclass[aps,prd,12pt,preprint,tightenlines,superscriptaddress,
amsfonts,amssymb,amsmath,byrevtex,showpacs,nofootinbib]{revtex4}
\usepackage{graphics}
\usepackage{epsfig,subfigure,amsmath}
\newcounter{mnotecount}[section]
\renewcommand{\themnotecount}{\thesection.\arabic{mnotecount}}
\newcommand{\mnote}[1]
{\protect{\stepcounter{mnotecount}}$^{\mbox{\footnotesize  $
      \bullet$\themnotecount}}$ \marginpar{\raggedright\tiny
    $\!\!\!\!\!\!\,\bullet$\themnotecount: #1} }

\begin{document}
\newcommand{\dR}{\mathbb R}
\newcommand{\dC}{\mathbb C}
\newcommand{\dZ}{\mathbb Z}
\newcommand{\dN}{\mathbb N}
\newcommand{\id}{\mathbb I}

\title{Turning big bang into big bounce: II. Quantum dynamics}

\author{Przemys{\l}aw Ma{\l}kiewicz$^\dag$ and
W{\l}odzimierz Piechocki$^\ddag$
\\ Theoretical Physics Department, Institute for Nuclear Studies
\\ Ho\.{z}a 69, 00-681 Warsaw, Poland;
\\ $^\dag$pmalk@fuw.edu.pl, $^\ddag$piech@fuw.edu.pl}

\date{\today}

\begin{abstract}
We analyze the  big bounce   transition of the quantum FRW model
in the setting of the nonstandard loop  quantum cosmology (LQC).
Elementary observables are used to quantize composite observables.
The spectrum of the energy density operator is bounded and
continuous. The spectrum of the volume operator is bounded from
below and discrete. It has equally distant levels defining a
quantum of the volume. The discreteness may imply a foamy
structure of spacetime at semiclassical level which may be
detected in astro-cosmo observations. The nonstandard LQC method
has a free parameter that should be fixed in some way to specify
the big bounce transition.
\end{abstract}
\pacs{98.80.Qc,04.60.Pp,04.20.Jb} \maketitle

\section{Introduction}
It is commonly expected that the cosmological singularity (CS)
problem \cite{MTW,SWH,JPAK,JMMS} may be resolved  in a theory
which unifies gravity and quantum physics. It seems that recent
developments concerning quantization of cosmological models  by
making use of  {\it loop geometry} may bring solution to the
problem. It consists in turning the  classical {\it big bang}
singularity into  {\it big bounce}, BB, transition. There exist
two methods to address the issue: {\it standard} loop quantum
cosmology (LQC) and {\it nonstandard} LQC. The former has been
developed during the last decade (see
\cite{Ashtekar:2003hd,Bojowald:2006da} and references therein) and
has been inspired by the loop quantum gravity, LQG (see
\cite{TT,CR,Ashtekar:2004eh} and references therein). The latter
has been proposed recently
\cite{Dzierzak:2008dy,Malkiewicz:2009zd,Dzierzak:2009ip,Dzierzak:2009dj}
and seems to be related to the reduced phase space quantization of
LQG \cite{Giesel:2007wn}.

The standard LQC means basically the Dirac  method of
quantization, which begins with quantization of the {\it
kinematical} phase space followed by imposition of constraints of
the gravitational system in the form of operators at the quantum
level. Finding kernels of these operators helps to define the {\it
physical} Hilbert space. In the nonstandard LQC (our method) one
first solves all the constraints at the classical level to
identify the {\it physical} phase space. Next, one identifies the
algebra of {\it elementary} observables (in the physical phase
space) and finds its representation. Then, {\it composite}
observables are expressed in terms of elementary ones and
quantized. Final goal is finding {\it spectra} of composite
observables which are used to examine the {\it nature} of the BB
phase in the evolution of the universe.

In what follows we restrict our considerations to  the  flat
Friedmann-Robertson-Walker (FRW) model with massless scalar field.
This model of the universe includes the initial cosmological
singularity and has been intensively studied recently within the
standard LQC.

In our recent paper \cite{Dzierzak:2008dy} we have argued  that
the resolution of the singularity offered by LQC requires specific
value of the fundamental {\it length}.  The size of this length
has not been determined satisfactory yet.  Present paper is an
extended version of \cite{Malkiewicz:2009zd} and together with the
classical formalism \cite{Dzierzak:2009ip} specifies our
nonstandard LQC (when applied to FRW model).

In order to have our paper self-contained,  we recall in Sec. II
some aspects of the classical formalism of our nonstandard LQC
\cite{Dzierzak:2009ip}. In Sec. III we present our quantization
procedure. It consists in  finding representation for composite
observables like the energy density and the volume function, and
calculating their spectra. We conclude in the last section.

In the Appendix, we give an extended {\it motivation} for our
paper. In particular, we justify the need for quantization of our
loop FRW model [12], which is free from a cosmological singularity
already at the classical level due to the {\it modification} of
the classical Hamiltonian.

\section{Classical Level}

\subsection{Hamiltonian}

The gravitational part of the classical Hamiltonian, $H_g$, of the
flat FRW model may be presented in the form \cite{Dzierzak:2009ip}
\begin{equation}\label{hamR}
    H_g = \lim_{\mu\rightarrow \,0}\; H^{(\mu)}_g ,
\end{equation}
where
\begin{equation}\label{hamL}
H^{(\mu)}_g = - \frac{sgn(p)}{2\pi G \gamma^3 \mu^3}
\sum_{ijk}\,N\, \varepsilon^{ijk}\, Tr \Big(h^{(\mu)}_i
h^{(\mu)}_j (h^{(\mu)}_i)^{-1} (h^{(\mu)}_j)^{-1}
h_k^{(\mu)}\{(h_k^{(\mu)})^{-1},V\}\Big) ,
\end{equation}
and where $V= |p|^{\frac{3}{2}}= a^3 V_0$ is the volume of the
elementary cell $\mathcal V\subset \Sigma$ ($\Sigma$ is spacelike
hyper-surface); $p$ is a canonical variable; the metric of $k=0$
FRW model is: $ds^2=-N^2(t)\,dt^2+a^2(t)\,(dx^2+dy^2+dz^2),$ where
$a$ is the scale factor and $N$ denotes the lapse function;
$\varepsilon^{ijk}$ is the alternating tensor;  $\gamma$ is the
Barbero-Immirzi parameter, and the holonomy function
$h_k^{(\mu)}$, along straight line of coordinate length
proportional to $\mu/|a|$,  reads
\begin{equation}\label{hol}
h^{(\mu)}_k (\beta) = \cos (\mu \beta/2)\;\id + 2\,\sin (\mu
\beta/2)\;\tau_k,
\end{equation}
where $\tau_k = -i \sigma_k/2\;$ ($\sigma_k$ are the Pauli spin
matrices).

The total Hamiltonian for FRW universe with a massless scalar
field, $\phi$, is found to be
\begin{equation}\label{ham}
   H = H_g + H_\phi ,
\end{equation}
where  $H_\phi = N\,p^2_\phi |p|^{-\frac{3}{2}}/2$, and where
$\phi$ and $p_\phi$ are canonical variables. Equation (\ref{ham})
satisfies the condition $ H\approx 0$ as it corresponds to the
{\it scalar constraint} of general relativity; the Gauss and
diffeomorphism constraints are equal zero (in the strong sense)
due to the choice of gauges.

Making use of (\ref{hol}) we calculate (\ref{hamL}) and get the
modified total Hamiltonian $H^{(\lambda)}_g$ corresponding to
(\ref{ham}) in the form \cite{Dzierzak:2009ip}
\begin{equation}\label{regH}
  H^{(\lambda)}/N = -\frac{3}{8\pi G \gamma^2}\;\frac{\sin^2(\lambda
\beta)}{\lambda^2}\;v + \frac{p_{\phi}^2}{2\, v} \approx 0,
\end{equation}
where
\begin{equation}\label{re1}
    \beta := \frac{c}{|p|^{1/2}},~~~v := |p|^{3/2}
\end{equation}
are the  canonical variables of the so called improved scheme, and
where $c$ and $p$ are canonical variables. The variable $\beta =
\gamma\dot{a}/a$ corresponds to the Hubble parameter $\dot{a}/a$,
and $v^{1/3} = a V_0^{1/3}$ is proportional to the scale factor
$a$. The relationship between the coordinate length $\mu$  (which
depends on $p$) and the physical length $\lambda$ (which is a
constant) reads:
    $\lambda = \mu \,|p|^{1/2} = \mu\, a \,V_0^{1/3}$.

At this stage, it should be emphasized that (\ref{regH}) presents
a purely classical Hamiltonian {\it modified} by the holonomy
(\ref{hol}), i.e. Eq. (\ref{regH}) includes no quantum physics.
Contrary, in the standard LQC (\ref{regH}) is called a
semi-classical or effective Hamiltonian and is interpreted to
include some `imprints' of quantization. In our nonstandard LQC
quantum physics enters the formalism only when quantizing the
algebra of observables (presented  in the subsequent section).

The complete Poisson bracket for the canonical variables
$(\beta,v,\phi,p_\phi)$ is defined to be
\begin{equation}\label{re2}
    \{\cdot,\cdot\}:= 4\pi G\gamma\;\bigg[ \frac{\partial \cdot}
    {\partial \beta} \frac{\partial \cdot}{\partial v} -
     \frac{\partial \cdot}{\partial v} \frac{\partial \cdot}{\partial \beta}\bigg] +
     \frac{\partial \cdot}{\partial \phi} \frac{\partial \cdot}{\partial p_\phi} -
     \frac{\partial \cdot}{\partial p_\phi} \frac{\partial \cdot}{\partial
     \phi}.
\end{equation}
The dynamics of a canonical variable $\xi$ is defined by
\begin{equation}\label{dyn}
    \dot{\xi} := \{\xi,H^{(\lambda)}\},~~~~~~\xi \in \{\beta,v,\phi,p_\phi\},
\end{equation}
where $\dot{\xi} := d\xi/d\tau$, and where $\tau$ is an evolution
parameter.

The dynamics in the  {\it physical} phase space,
$\mathcal{F}_{phys}^{(\lambda)}$, is defined by solutions to
(\ref{dyn}) satisfying the constraint $H^{(\lambda)}\approx 0$.
The solutions of (\ref{dyn}) ignoring the constraint  are in the
{\it kinematical} phase space.

\subsection{Elementary  observables}

A function, $\mathcal{O}$, defined on phase space   is a Dirac
{\it observable}  if
\begin{equation}\label{dirac}
\{\mathcal{O},H^{(\lambda)}\} \approx 0.
\end{equation}

Equation (\ref{dirac}) leads to \cite{Dzierzak:2009ip}
\begin{equation}\label{dir}
\frac{\sin(\lambda\beta)}{\lambda}\,\frac{\partial
\mathcal{O}}{\partial\beta} - v \cos(\lambda\beta)\,\frac{\partial
\mathcal{O}}{\partial v} - \frac{\kappa\,\textrm{sgn}(p_{\phi})}{4
\pi G}\,\frac{\partial \mathcal{O}}{\partial\phi} = 0,
\end{equation}
where $\kappa^2 := 4\pi G/3$.  Solution to (\ref{dir}) in
$\mathcal{F}_{phys}^{(\lambda)}$ is found to be
\cite{Dzierzak:2009ip}
\begin{equation}\label{obser1}
\mathcal{O}_1:= p_{\phi},~~~\mathcal{O}_2:= \phi -
\frac{\textrm{sgn}(p_\phi)}{3\kappa}\;\textrm{arth}\big(\cos(\lambda
\beta)\big).
\end{equation}
One may parameterize $\mathcal{F}_{phys}^{(\lambda)}$ by the {\it
elementary} observables $\mathcal{O}_1$ and $\mathcal{O}_2$, and
define the Poisson structure as follows \cite{Dzierzak:2009ip}
\begin{equation}\label{sym4}
\{\mathcal{O}_2,\mathcal{O}_1\}=
1,~~~~~~\{\cdot,\cdot\}:=\frac{\partial\cdot}{\partial
\mathcal{O}_2}\frac{\partial\cdot}{\partial \mathcal{O}_1} -
\frac{\partial\cdot}{\partial
\mathcal{O}_1}\frac{\partial\cdot}{\partial \mathcal{O}_2}.
\end{equation}
For simplicity we use the same notation for the Poisson bracket in
(\ref{sym4}) and  (\ref{re2}).

\subsection{Compound observables}

In what follows we consider functions on {\it physical} phase
space like the energy density and the volume, which describe
singularity aspects of our cosmological model. Considered
functions are expressed in terms of elementary observables and an
evolution parameter $\phi$. They become observables for each fixed
value of $\phi$, since in such case they are only functions of
observables \cite{Dzierzak:2009ip}.

An expression for the energy density $\rho$ of the scalar field
$\phi$ reads \cite{Dzierzak:2009ip}
\begin{equation}\label{rho2}
\rho(\phi,\lambda)=\frac{1}{2}\,\frac{1}{(\kappa\gamma\lambda)^2\,
    \cosh^2 3\kappa  (\phi- \mathcal{O}_2)}.
\end{equation}
The volume, a geometrical function, is found to be
\cite{Dzierzak:2009ip}
\begin{equation}\label{vol}
    v(\phi,\lambda) = \kappa\gamma\lambda\,
    |\mathcal{O}_1|\,\cosh3\kappa  (\phi-
    \mathcal{O}_2).
\end{equation}

\section{Quantization}

By quantization we mean: (i) finding a self-adjoint representation
of composite observables, and (ii) calculating spectra of
operators corresponding to the composite observables.

\subsection{Representation of elementary observables}

In what follows we use two representations  of the classical
algebra (\ref{sym4}). Namely,
\begin{equation}\label{quant1}
\mathcal{O}_1 \longrightarrow \widehat{\mathcal{O}}_1 f(x):=
-i\,\hbar\,\partial_x f(x),~~~~\mathcal{O}_2 \longrightarrow
\widehat{\mathcal{O}}_2 f(x):= \widehat{x} f(x) := x f(x),
\end{equation}
which leads to $[\widehat{\mathcal{O}}_1,\widehat{\mathcal{O}}_2]=
- i\,\hbar\,\id$, and
\begin{equation}\label{quant2}
\mathcal{O}_1 \longrightarrow \widehat{\mathcal{O}}_1 f(x):=
\widehat{x} f(x) := x f(x),~~~~\mathcal{O}_2 \longrightarrow
\widehat{\mathcal{O}}_2 f(x):= -i\,\hbar\,\partial_x f(x),
\end{equation}
that gives $[\widehat{\mathcal{O}}_1,\widehat{\mathcal{O}}_2]=
i\,\hbar\,\id$, where $x\in \mathbb{R}$. Due to the Stone$-$von
Neumann theorem all self-adjoint representations of the algebra
(\ref{sym4}) are unitarily equivalent to the representation
(\ref{quant1}) or (\ref{quant2}) defined on a suitable dense
subspace of $L^2(\dR)$. In this sense our choice of representation
for (\ref{sym4}) is unique.

\subsection{Energy density}

The representation (\ref{quant2}) is  essentially self-adjoint on
on the dense subspace $D$ of the Hilbert space $L^2 [-r,r],\,$
where $r \in \dR_+$, defined to be
\begin{equation}\label{den}
D:= \{f\in C^\infty [-r,r]\,|\, f^{(n)}(-r) = f^{(n)}(r),\, n \in
\{0\}\cup \dN \},
\end{equation}
where $f^{(n)} := d^{\,n}f/dx^n$. The eigenvalue problem,
$\widehat{\mathcal{O}}_2 f_p = p\, f_p$, has the solution
\begin{equation}\label{sol1}
f_p (x) = (2r)^{-1/2} \exp(i x p /\hbar),~~~~p(k):= 2\pi\hbar
k/r,~~k \in \dZ .
\end{equation}
The  spacing $ \square$ of neighboring eigenvalues
\begin{equation}\label{spa}
\square := p(k+1) - p(k) = 2\pi\hbar/r
\end{equation}
can be made as small as desired by making $r$ sufficiently large.
Thus, one may say that the  spectrum of $\widehat{\mathcal{O}}_2 $
is  continuous.

In the representation (\ref{quant2}) the energy density operator
reads
\begin{equation}\label{quant3}
 \widehat{\rho}:=\frac{1}{2}
    \frac{1}{(\kappa\gamma\lambda)^2\cosh^2 3\kappa(\phi+i\,\hbar\,\partial_x)}.
\end{equation}
Since $\widehat{\mathcal{O}}_2$ is {\it essentially} self-adjoint
on $\mathcal{F}_r := \{f_{p(k)}\}_{k \in \dZ}$, we may apply the
spectral theorem to get
\begin{equation}\label{quant4}
\widehat{\rho}\,f_p = \rho (\phi,\lambda,p)\,f_p ,
\end{equation}
where
\begin{equation}\label{quant5}
\rho (\phi,\lambda,p) := \frac{1}{2} \frac{1}{(\kappa\gamma
\lambda)^2\cosh^2 3\kappa(\phi - p)},
\end{equation}
and where $\rho (\phi,\lambda,p)$ is the eigenvalue corresponding
to the eigenvector $f_p$.

Our results mean that classical (\ref{rho2}) and quantum
(\ref{quant5}) expressions for the energy density {\it coincide}.
It is clear that the maximum density, $\rho_{max}$, reads
\begin{equation}\label{maxio}
\rho_{max}(\lambda) = \frac{1}{2}\frac{1}{(\kappa\gamma
\lambda)^2}.
\end{equation}
Starting from the representation (\ref{quant1}), instead of
(\ref{quant2}), we would get the quantum model of the energy
density presented in \cite{Malkiewicz:2009zd}, which is equivalent
to the present one.

\subsection{Volume operator}

To define the quantum operator corresponding to $v$, we use
\begin{equation}\label{vol1}
v = |w|,~~~w :=
\kappa\gamma\lambda\;\mathcal{O}_1\;\cosh3\kappa(\phi-
\mathcal{O}_2).
\end{equation}
Thus, quantization of $v$ reduces to the  quantization problem of
$w$. Quantization of the latter may be done in a standard way as
follows
\begin{equation}\label{vol2}
\hat{w}\,f(x) :=
    \kappa\gamma\lambda\,\frac{1}{2}\,\bigg(
    \widehat{\mathcal{O}}_1\,\cosh3\kappa  (\phi-
    \widehat{\mathcal{O}}_2)+
     \cosh3\kappa  (\phi-
    \widehat{\mathcal{O}}_2)\;\widehat{\mathcal{O}}_1\bigg) f(x),
\end{equation}
where $f \in  L^2 (\dR)$. For the elementary observables
$\mathcal{O}_1$ and $\mathcal{O}_2$ we use the  representation
(\ref{quant1}). An explicit form of the operator $\hat{w}$ is
\begin{equation}\label{repp1}
\hat{w}\,f(x)= i\,\frac{\kappa\gamma\lambda\hbar}{2}\bigg(
    2 \cosh3\kappa(\phi-x)\;\frac{d}{dx}
     -3\kappa\sinh3\kappa
    (\phi-x)\bigg)\,f(x).
\end{equation}
Considerations simplify if we take $f$ in the form
\begin{equation}\label{eq3}
f(x) :=  A\,e^{i h(x)}\, \cosh^{-1/2} 3 \kappa (\phi - x),
\end{equation}
where $A \in \dR $, and where $h$ is a real-valued function.

\subsubsection{Eigenvalue problem}

Let us consider the eigenvalue problem for the operator $\hat{w}$
in the set of functions of the form (\ref{eq3}). We have
\begin{equation}\label{eq4}
\hat{w}\, f (x)= -\kappa\gamma\lambda\hbar
\cosh3\kappa(\phi-x)\frac{d h(x)}{dx}\, f(x) =: b\, f (x),
\end{equation}
where  $b \in \dR $ is the eigenvalue of $\hat{w}$. One may verify
that a general form of $h$ satisfying (\ref{eq4}) is given by
\begin{equation}\label{eq55}
 h (x)=  \frac{2  b}{3\kappa^2
    \gamma\lambda\hbar}\arctan
    e^{3\kappa(\phi-x)}.
\end{equation}
Thus, a normalized $f_b $ satisfying (\ref{eq4}) reads
\begin{equation}\label{eq5}
f_b (x):= \frac{\sqrt{\frac{3\kappa}{\pi}}\exp\Big(i \frac{2
b}{3\kappa^2 \gamma\lambda\hbar}\arctan
    e^{3\kappa(\phi-x)}\Big)}{\cosh^{\frac{1}{2}}3\kappa(\phi-x)}.
\end{equation}

\subsubsection{Orthogonality}

Making use of (\ref{eq5}) gives
\begin{equation}
    \langle f_b|f_a\rangle= \frac{3\kappa}{\pi}\int_{-\infty}^{\infty}\frac{\exp\Big(i \frac{2
(a-b)}{3\kappa^2 \gamma\lambda\hbar}\arctan
    e^{3\kappa(\phi-x)}\Big)}{\cosh3\kappa(\phi-x)}~dx .
\end{equation}
We introduce a new variable $y=e^{3\kappa(\phi-x)}$ and have
\begin{equation}
   \langle f_b|f_a\rangle=\frac{2}{\pi}\int_{0}^{\infty}\frac{\exp\Big(i \frac{2
(a-b)}{3\kappa^2 \gamma\lambda\hbar}\arctan
    y\Big)}{1+y^2}~dy .
\end{equation}
Another substitution $\tan z =y$ leads to
\begin{equation}
   \langle f_b|f_a\rangle=\frac{2}{\pi}\int_{0}^{\frac{\pi}{2}}\exp\Big(i \frac{2
(a-b)}{3\kappa^2 \gamma\lambda\hbar}z\Big)~dz=-i\,
\frac{3\kappa^2\gamma\lambda\hbar}{\pi(a-b)}\exp\Big(i \,\frac{2
(a-b)}{3\kappa^2 \gamma\lambda\hbar}z\Big)\Big|_0^{\frac{\pi}{2}}
.
\end{equation}
It is clear that $\langle f_b|f_a\rangle = 0\;$ iff
\begin{equation}\label{ab}
    a-b = 6\kappa^2\gamma\lambda\hbar\,m = 8\pi
    G\gamma\lambda\hbar\,m,~~~~m\in \mathbb{Z} .
\end{equation}
Thus, the set $\mathcal{F}_b:=\{~f_a\;|\; a = b + 8\pi
G\gamma\lambda\hbar\,m;~m\in \dZ;~b \in \dR~\}$ is {\it
orthonormal}. Each subspace $\mathcal{F}_b \subset L^2(\dR)$ spans
a pre-Hilbert space.  The completion of each $span
\;\mathcal{F}_b,~~\forall b \in \dR$, in the norm of $L^2(\dR)$
gives $L^2(\dR)$.

\subsubsection{Self-adjointness}

Since $\langle f_b|\hat{w}f_a\rangle - \langle
\hat{w}f_b|f_a\rangle = (a-b)\langle f_b|f_a\rangle$, the operator
$\hat{w}$ is {\it symmetric} on $\mathcal{F}_b $ for any $b \in
\dR$, because $\langle f_b|f_a\rangle=0$ for $a\neq b$ due to the
orthogonality of the set $\mathcal{F}_b$.

To examine the self-adjointness of the unbounded operator
$\hat{w}$, we first identify the deficiency subspaces,
$\mathcal{K_\pm}$,  of this operator \cite{RS,DS}
\begin{equation}\label{self1}
\mathcal{K_\pm}:= \{g_\pm \in D_b (\hat{w}^\ast)~|~\langle
g_\pm|(\hat{w} \pm i \id)f_a \rangle =0, \;\forall f_a \in D_b
(\hat{w})\},
\end{equation}
where $D_b (\hat{w}):=$ {\it span} $\mathcal{F}_b$, and where
$D_b(\hat{w}^{\ast}):=\{f\in L^2(\dR) :~\exists !f^{\ast}~\langle
f^{\ast}|g\rangle=\langle f|\hat{w}g\rangle,~\forall g\in
D_b(\hat{w})\}$.

For each $f_a \in D_b(\hat{w})\subset L^2(\dR)$ we have
\begin{equation}\label{self2}
0 = \langle g_\pm|(\hat{w} \pm i \id)f_a\rangle = (a \pm i)
\int_{-\infty}^\infty dx \;\overline{g_\pm (x)} f_a
(x)~~~~~\Longrightarrow~~~~~g_+ = 0 = g_- .
\end{equation}
Thus, the deficiency indices $n_\pm := dim [\mathcal{K}_\pm]$ of
$\hat{w}$ satisfy the relation: $n_+ = 0 = n_-$, which proves that
the operator $\hat{w}$ is {\it essentially} self-adjoint on
$D_b(\hat{w})$.

\subsubsection{Spectrum}

Due to the spectral theorem on self-adjoint operators
\cite{RS,DS}, we may carry out quantization of the volume as
follows
\begin{equation}\label{sp1}
 v = |w|~~~\longrightarrow~~~\hat{v} f_a :=  |a| f_a .
\end{equation}

A common feature of all $\mathcal{F}_b$ is the existence of the
minimum gap $\bigtriangleup  := 8\pi G\gamma\hbar\,\lambda\;$ in
the spectrum, which defines a  {\it quantum} of the volume.  Let
us discuss this issue in more detail. Let us denote the minimum
eigenvalue of $\hat{v}$ by $v_{min}$. One can verify that
$v_{min}=\textrm{min}\{b,\Delta-b\}$, where $b\in [0,\Delta[$. The
spectrum consists of the union of $\{v_{min}+n\Delta\}$ and
$\{-v_{min}+(n+1)\Delta\}$, where $n=0,1,\dots$ Only in two cases
these two subsets  are identical, namely when $\,v_{min}=0\,$ or
$\,v_{min}=\Delta/2\,$, for which the minimum gap $\Delta$ is a
{\it constant} gap between any two adjacent levels of the
spectrum. Otherwise, the gap  equals either $\,\Delta -
2v_{min}\,$ or $\,2v_{min}\,$, and the minimum gap is the smaller
one. One can verify that the case of any $b \in \dR$ reduces to
the above case.

In the limit $\lambda \rightarrow 0$, corresponding to the
classical FRW model without the loop geometry modification, there
is no quantum of the volume.

 It results from (\ref{ab}) that for $b=0$ and $m=0$  the
minimum eigenvalue of $\hat{v}$ equals zero. It is a special case
that corresponds to the classical situation when $v=0$, which due
to (\ref{regH}) means that $p_\phi = 0$ so there is no classical
dynamics (for more details see \cite{Dzierzak:2009ip}). Thus, we
have a direct correspondence between classical and quantum levels
corresponding to this very special state. It is clear that all
other states describe  bouncing {\it dynamics}.

As the universe expands, a discrete spectrum of the volume
operator may favour a  {\it foamy} like structure which should
finally turn into a continuous spacetime. The quantum of a volume
may be used as a measure of a size, $\lambda_f$, of a spacetime
foam. One may speculate that $\lambda_f := \bigtriangleup^{1/3} =
\big(8\pi G\gamma\hbar\,\lambda\big)^{1/3}$.  Thus, an astro-cosmo
data that determine a size of  spacetime granularity may fix the
minimum length parameter $\lambda$ of LQC. That would enable
making an estimate of the critical matter density $\rho_{\max} =
1/2(\kappa \gamma \lambda)^2$ corresponding to the Big Bounce.

The granularity of volume should lead to the granularity of energy
of physical fields. We suggest, making use of the de Broglie
relation, that a specific  particle representing a quantum of
energy may have a momentum $p_i$ corresponding to its wavelength
$\lambda_i$ such that $p_i\,\lambda_i = \hbar$. The detection of
an ultrahigh energy particle with specific $p_i$ may be used to
determine $\lambda_i$, and consequently set the upper limit for
the fundamental length $\lambda_f$. The set of parameters
$\lambda_i$ (for a set of particles) may be treated further as
multiplicities of $\lambda_f$ in which case the greatest common
divisor of all $\lambda_i$ would set the lowest upper limit for
$\lambda_f$.

\subsubsection{Evolution}

The relation between eigenvectors corresponding to the same
eigenvalue for different values of the parameter $\phi$ reads:
\begin{equation}\label{tay1}
f_a^{\phi+\psi} = e^{\psi\partial_\phi}f_a^{\phi} =
e^{-i\frac{\psi}{\hbar} \widehat{\mathcal{O}}_1}f_a^{\phi} .
\end{equation}

One may verify that
\begin{equation}\label{da1}
    \hat{w}(\phi+\psi)=\cosh{(3\kappa\psi)}\hat{w}(\phi)
    +\frac{\sinh{(3\kappa\psi)}}
    {3\kappa}\partial_{\phi}\hat{w}(\phi),
\end{equation}
thus
\begin{eqnarray}\nonumber
    &&\langle f_b^{\phi}|\hat{w}(\phi+\psi)f_a^{\phi}\rangle =
    \langle f_b^{\phi}|\hat{w}(\phi)f_a^{\phi}\rangle\cosh{(3\kappa\psi)}+
    \frac{\sinh{(3\kappa\psi)}}{3\kappa}\langle f_b^{\phi}|\partial_{\phi}
    \hat{w}(\phi)f_a^{\phi}\rangle\\&&=a\cosh{(3\kappa\psi)}\,
    \delta_{ab}+(b-a)\frac{\sinh{(3\kappa\psi)}}{3\kappa}\langle
    f_b^{\phi}|\partial_xf_a^{\phi}\rangle .
\end{eqnarray}
Finally, an evolution of the  expectation value of the operator
$\hat{w}$ is found to be
\begin{equation}\label{da2}
    \langle
    f (\phi)|\hat{w}(\phi+\psi)f(\phi)\rangle = A \cosh3\kappa(\psi+B),
\end{equation}
where $f:=\sum\alpha_a f_a, \; f_a \in \mathcal{F}_b$.   One may
verify that
\begin{equation}
    A=\textrm{sgn}(X)\sqrt{X^2-Y^2},~~~~B=\frac{1}{6\kappa}\ln{\frac{X+Y}{X-Y}}~,
\end{equation}
where
\begin{equation}
    X:=\sum_{a}|\alpha_a|^2a,~~~~Y:=\sum_{a,~m}\frac{\bar{\alpha}_b\alpha_a-
    \bar{\alpha}_a\alpha_b}
   {i\pi}\frac{m(2a+6m\kappa^2\gamma\hbar\lambda)}{(2m-1)(2m+1)},
\end{equation}
and where $b = a + 6\kappa^2\gamma\lambda\hbar, \;\; b\in \dR, \;
~m\in \mathbb{Z}$, and $|X|>|Y|$.

We can see that the evolution of the expectation value of the
operator $\hat{w}$ coincides with the classical expression
(\ref{vol1}).

\subsection{Energy density operator in the basis $\mathcal{F}_b$}

The operator $\hat{\rho}$ may be expressed in terms of the basis
vectors from $\mathcal{F}_b$. One may verify that
\begin{eqnarray}
\langle f_{a+m \bigtriangleup}|\hat{\rho} f_{a}\rangle= \left\{
\begin{array}{lll}
         \frac{-1}{8(\kappa\gamma\lambda)^2} & \mbox{if $\;\;m = 1$ or $-1$};\\
        \frac{1}{4(\kappa\gamma\lambda)^2} & \mbox{if $\;\;m = 0$};\\
        0 & \mbox{otherwise}\end{array} \right.
\end{eqnarray}
which leads to
\begin{equation}\label{nice}
\hat{\rho} f_{a}=
\frac{1}{4(\kappa\gamma\lambda)^2}f_{a}-\frac{1}{8(\kappa\gamma\lambda)^2}
f_{a-4 \bigtriangleup}-\frac{1}{8(\kappa\gamma\lambda)^2}f_{a+4
\bigtriangleup} ,
\end{equation}
where $\bigtriangleup  := 8\pi G\gamma\hbar\,\lambda\;$.
Therefore, $\hat{\rho}$ is bounded and does not commute with
$\hat{v}$. The latter is consistent with the Poisson bracket
relation for these observables, which reads
\begin{equation}\label{nonco}
\{w,\rho\}=  3 \kappa^2 \gamma \lambda \sinh 3\kappa  (\phi-
\mathcal{O}_2)\,\rho.
\end{equation}
Therefore, the operators $\hat{\rho}$ and $\hat{v}$ cannot have
common eigenfunctions so they provide two alternative ways of
describing the system, either in terms of the eigenfunctions of
$\hat{\rho}$ or $\hat{v}$.

\section{Summary and Conclusions}

Turning the big bang into the big bounce (BB) in the FRW universe
is due to the modification of the model at the {\it classical}
level by making use of the loop geometry. The modification is
parameterized by a {\it continuous} parameter  $\lambda $.

Each value of $\lambda$ specifies the critical energy density of
the scalar field corresponding to the BB. The spectrum of the
energy density operator is {\it bounded} and {\it continuous}.
Classical and quantum expressions for the minimum of the energy
density coincide. The spectrum of the volume operator,
parameterized by $\lambda$, is {\it bounded} from below and {\it
discrete}.  The expectation value of the volume operator coincide
with the classical expression. The results concerning the volume
operator may be extended to the area and the length operators
\cite{Dzierzak:2009ip}.

An {\it evolution} parameter, $\phi$, does not belong to the {\it
physical} phase space of our nonstandard LQC. Thus, it stays
classical during the quantization process
\cite{Malkiewicz:2009zd}. In the standard LQC, contrary to our
method, $\phi$ is a phase space variable so it must be quantized
\cite{Ashtekar:2006wn}. Being a quantum variable it may {\it
fluctuate} so its use as an evolution parameter at the quantum
level has poor interpretation.

The Hamiltonin {\it constraints} are treated differently in the
standard and nonstandard LQC at the classical level. In the former
case the Gauss and diffeomorphism constraints are turned into zero
by a suitable choice of gauges; the scalar constraint is solved
only at the quantum level.  In the latter case all the constraints
are classically solved leading to the {\it physical} phase space.
Thus, considered observables are  not kinematical, but physical.
This is why an evolution of geometrical functions like the volume
may be used for testing the singularity aspects of a given
cosmological model. The standard and our LQC methods give similar
results in the sense that they lead to  the Big Bounce transition.
However, our method is fully controlled analytically as it does
not require any numerical work.

We believe that our nonstandard LQC may be related with the
reduced phase space quantization of Loop Quantum Gravity (LQG)
(see \cite{Giesel:2007wn} and references therein). Finding the
correspondence may help  deriving LQC from LQG.

There exist results concerning the spectrum of the volume operator
obtained within LQG (see, e.g.
\cite{Ashtekar:1997fb,Meissner:2005mx}), but cannot be compared
easily with our results due to the lack of a direct correspondence
between LQG and LQC models.

As there is no specific choice of $\lambda$, the BB may occur at
any low and high densities \cite{Dzierzak:2008dy}. Finding
specific value of the parameter $\lambda$ (and energy scale
specific to the BB) is an open problem. It may happen, that the
value of the parameter $\lambda$ cannot be  determined, for some
reason, theoretically. Fortunately, there is a rapidly growing
number of data coming from observational cosmology that may be
useful in this context. In particular, the detection of the
primordial gravitational waves created at  the BB phase may bring
valuable information about this phase
\cite{Mielczarek:2008pf,Calcagni:2008ig,Grain:2009kw}. There
exists speculation that the foamy structure of spacetime may lead
to the dependence of the velocity of a photon on its energy. Such
dependance is weak, but may sum up to give a measurable effect in
the case of  photons travelling over cosmological distances across
the Universe \cite{GAC}. Presently, available data suggest that
such {\it dispersion} effects do not occur up to the energy scale
$5\times 10^{17}$ GeV \cite{Aharonian:2008kz}  so such effects may
be present, but at higher energies.

Various forms of  discreteness of spacetime underly many
approaches in fundamental physics. Just to name a few:
noncommutative geometry \cite{Heller:2006uq}, causal sets approach
\cite{Rideout:2008df}, gravitational Wilson loops
\cite{Hamber:2009uz}, Regge calculus \cite{Bahr:2009qc}, path
integral over geometries \cite{Ambjorn:2009ts}, spin foam model
\cite{Kaminski:2009fm}, and categories \cite{Baez:2009as}. The
discreteness may translate at the semi-classical level into a
foamy structure of space. Such expected property of spacetime
creates large activity in observational astrophysics and cosmology
(see, e.g. Lorentz and CPT violation \cite{Kostelecky:2008ts},
dispersion of cosmic photons \cite{AmelinoCamelia:2009pg},
electrons \cite{Galaverni:2008yj} and neutrinos
\cite{Ellis:2008fc}, birefringence effects \cite{Gleiser:2001rm}).
Our results concerning the physics of geometry at short distances
give some support to these approaches and expectations.

For more concluding remarks, connected with motivation for our
approach, we recommend reading the appendix section.

\begin{acknowledgments}
We are grateful to Piotr Dzier\.{z}ak for helpful discussions.
\end{acknowledgments}

\appendix
\section{Motivation}

In what follows, we give an extended {\it motivation} for our
paper. In particular, we justify the need for quantization of our
loop FRW model \cite{Dzierzak:2009ip}, which is free from the
cosmological singularity already at the classical level due to the
{\it
modification} of the classical Hamiltonian:\\

If we intend to quantize canonically  a system with Hamiltonian
{\it constraint}, we can apply two methods: (i) Dirac's
quantization - quantize  kinematics ignoring the constraint, than
impose the constraint at the quantum level (scheme D), or (ii)
reduced phase space quantization - solve the constraint at
classical level, than quantize the system free of the constraint
(scheme R).

Suppose there is a {\it theoretical} reason to quantize a
classical system with  the Hamiltonian constraint, but
experimental/observational data with expected {\it quantum}
effects are not available yet. This is the situation specific for
gravitational systems. Theoretical reasons to quantize gravity
are, e.g., singularities of classical theory and belief that all
interactions can be unified. Which method should we choose to
achieve the goal?  Since the quantum data are not available yet,
the best strategy seems to be applying both methods to {\it
compare} the results. An agreement of the results would prove that
the procedure of quantization was correct. Final test would be an
agreement of theoretical predictions with real world, i.e.
experimental/observational data (when the data become available).
An agreement at this stage would end the story.

Standard LQC and LQG means the scheme D (DS), nonstandard LQC (our
method) corresponds to the scheme R (RS). All results (with small
exceptions) have been obtained so far within DS: almost 25-years
of efforts (hundreds of papers) in the case of LQG, and almost
10-years of efforts (hundreds of papers) in the case of LQC. There
is almost no criticism of these results (with small exception:
hep-th/0601129, hep-th/0501114).

The above strategy underlies our paper: we intend to
verify/confirm the DS LQC  results.  On the other hand, an
alternative method may broaden our understanding of various
conceptual issues like, e.g., quantum evolution of a system with
the Hamiltonian constraint or defining physical observables. Now,
let us be more specific. The technical procedure (with different
interpretation) leading up to Eq. (\ref{regH}) (presenting a
Hamiltonian parametrized by $\lambda$) is identical in both DS and
RS. The real difference arises when one begins with the {\it
implementation} of the Hamiltonian constraint.

In DS, one promotes Eq. (\ref{regH}) to an operator equation at
the {\it quantum} level (see, gr-qc/0304074, gr-qc/0604013,
gr-qc/0607039) with  the parameter $\lambda$ {\it different} from
zero\footnote{For simplicity, we ignore the difference between the
so called `old' and `improved' schemes notations and denote the
regularization parameter by the same symbol $\lambda$.}. The
question arises: Why is the parameter $\lambda$ kept {\it
different} from zero?  The technical answer is: In the limit
$\lambda \rightarrow 0 $ the {\it regular} constraint equation
turns into the {\it singular} Wheeler-DeWitt equation. Users of DS
propose the following justification for keeping $\lambda \neq 0$:
In the {\it loop} representation used to quantize the kinematical
level (first step of DS) this representation does not exist  for
$\lambda = 0$. In the second step of DS, one turns the classical
constraint equation into an operator equation, which one defines
consequently in the {\it loop} representation of the kinematical
level. This is why one keeps $\lambda \neq 0$ in this operator
equation. This procedure provokes the next question: Why one uses
such an {\it exotic} representation which is not defined for
$\lambda \neq 0 ?$ The explanation of the users of DS is the
following: such representation is an analog of the representation
used at the kinematical level of loop quantum {\it gravity} (LQG).
Is it a satisfactory justification? Well, LQG has not been
constructed yet: the representation of the constraints algebra,
based on the achievements of kinematical level, has not been
found. The problem is extremely difficult because the algebra is
not a Lie, but a Poisson algebra (one structure function is not a
constant but a function on phase space). The users of the DS
method  deeply believe that sooner or later this problem of LQG
will be solved and LQC will be derived from LQG, so making use of
an analogy to LQG is a healthy approach.    This is a highly
promising development, but far from being completed.

In the RS case, the point of departure (as we said above) is Eq.
(\ref{regH}). The Hamiltonian must be the same, in  DS and RS,
otherwise the comparison of both methods would be impossible. Let
us emphasize that Eq. (\ref{regH})  {\it is not} an effective
semi-classical Hamiltonian in the RS method. It is considered to
be a purely {\it classical}, but {\it modified} Hamiltonian.  One
can treat it as a one-parameter family of classical Hamiltonians,
including the usual general relativity Hamiltonian as a special
case for $\lambda = 0$.

It can be seen easily that the singularity becomes `resolved' at
the classical level, for $\lambda \neq 0$, due to the functional
form of Eq. (\ref{regH}): big bang turns into big bounce. Why
should we quantize a cosmological model which is free from the
cosmological singularity?

\noindent There are at least three reasons:

\noindent (i) to make comparison with DS results, we must have a
quantum model;

\noindent (ii) the parameter $\lambda$ specifying the modification
is a {\it free} parameter in RS.  As the result, the classical
critical density of matter at the bounce becomes unspecified as it
depends on $\lambda$ (see, Eq. (\ref{rho2})). Since it may become
arbitrarily big for small enough $\lambda$, the system may enter
an arbitrarily small length scale, where quantum effects cannot be
ignored. We have discussed this issue with all details in
\cite{Dzierzak:2009ip} (subsection A of section V);

\noindent (iii) making predictions of our model for quantum cosmic
data may be used to {\it fix} the free parameter $\lambda$, after
such data become available.

Since the {\it quantum} energy density operator turns out to have
eigenvalues coinciding with the classical expression for the
energy density (see, Eqs. (\ref{rho2})  and (\ref{quant5})), no
new opportunity exists to fix $\lambda$ in the context of energy
density. Fortunately, the {\it quantum} volume operator has {\it
discrete} spectrum depending explicitly on $\lambda$. Since this
results implies that the space may have a {\it foamy} like
structure, the determination of $\lambda$ may be possible (see,
Sec. IV). Turning the argument around, the detection of such
effects could be used to argue that quantum gravity effects do
{\it occur} near the cosmological singularity.

The RS method reveals that the choice of the exotic Bohr
representation at the quantum kinematical level of DS {\it in
effect} corresponds to the setting $\lambda \neq 0$  at the
classical level.  After making standard quantization we keep
$\lambda \neq 0$ because as $\lambda \rightarrow 0$ the maximum
energy density blows up due to Eq. (\ref{maxio})), and we get a
singularity. Also the spectrum of the volume operator becomes {\it
continuous}, since $\bigtriangleup := 8\pi
G\gamma\hbar\,\lambda\;$ goes to zero as $\lambda \rightarrow 0$.
Thus, any volume of space may smoothly collapse to a singularity.
In both RS and DS methods, $\lambda$ cannot be treated as a {\it
regularization} parameter that one should remove from final
results. Quite the opposite, its presence is necessary for the
resolution of the singularity problem.

Our results agree with the results obtained within the DS method
in the sense that the classical initial singularity is {\it
replaced} with the big bounce transition.

In fact, we have done much more.   We have shown  that the
spectrum of the physical volume operator is {\it discrete}.  {\it
Physical} means that the spectrum may be used for comparison with
observational data, in contrast to the case of the spectrum of the
kinematical operators. We have also made some preliminary steps
towards the solution of the {\it evolution} problem, which is
fundamental for systems with Hamiltonian being the constraint of
the system.

We have found that LQC should be directly linked with {\it
experimental} data to become specific.  It is so because
$\lambda$, parametrizing the loop geometry, is a {\it free}
parameter of LQC. It seems, there is no satisfactory way to
determine it {\it theoretically} within LQC, but forthcoming
cosmic observations may solve the problem. We have already
addressed this issue in the context of the DS  (see, Sec 4 of
\cite{Dzierzak:2008dy}. Present paper confirms this result within
the RS approach. Some preliminary agreement with our results
(without reference to ours) may be found in an updated version of
the paper on the robustness of LQC: see, arXiv: 0710.3565, {\it
version 4}, Sec VI. One discusses there `parachuting by hand'
results from full LQG into LQC, since the derivation of LQC from
LQG has not been obtained yet. If we suitably fix $\lambda$ in our
method `by hand', we will get the numerical results concerning the
big bounce, like the critical energy density of matter field,
identical to the ones obtained within the DS method (see, Subsec A
of Sec V in \cite{Dzierzak:2009ip}).

Our method {\it relies} on a direct link with observational data
due to the unknown value of $\lambda$. This may make our model
suitable for describing observational data despite the fact that
FRW may have too much symmetry to be a realistic model of the
Universe. Taking theoretically determined $\lambda$ from {\it
incomplete} LQG, in DS method, seems to give a model of the
Universe less realistic than ours. Lacking of theoretically
determined numerical value of $\lambda$ is not drawback, but
advantage of the RS method.

\end{document}